\documentclass{article}

\usepackage{arXiv}

\usepackage[utf8]{inputenc} 
\usepackage[T1]{fontenc}    
\usepackage{hyperref}       
\usepackage{url}            
\usepackage{booktabs}       
\usepackage{amsfonts}       
\usepackage{nicefrac}       
\usepackage{microtype}      
\usepackage{lipsum}		
\usepackage{amsmath, amssymb}
\usepackage{multicol}
\usepackage{multirow}
\usepackage{caption}
\usepackage{xcolor}
\usepackage{blindtext}
\usepackage{graphicx}
\usepackage{doi}
\usepackage[affil-it]{authblk}
\usepackage{cleveref}
\crefname{figure}{Fig}{Figs}
\crefname{table}{Table}{Tables}
\usepackage{booktabs}
\usepackage{setspace} 
\usepackage{relsize}

\usepackage{xcolor}
\hypersetup{
    colorlinks,
    linkcolor={red},
    citecolor={blue},
    urlcolor={blue!80!black}
}

\title{Reducing Overtreatment of Indeterminate Thyroid Nodules Using a Multimodal Deep Learning Model}

\author[1]{Shreeram Athreya}
\author[2]{Andrew Melehy}
\author[3]{Sujit Silas Armstrong Suthahar}
\author[4]{Vedrana Ivezić}
\author[3]{Ashwath Radhachandran}
\author[5]{Vivek Sant}
\author[6]{Chace Moleta}
\author[4]{Henry Zheng}
\author[7]{Maitraya Patel}
\author[7]{Rinat Masamed}
\author[1,3,4,6,7]{Corey W. Arnold}
\author[3,4,7,$\ast$]{William Speier}

\affil[1]{Department of Electrical and Computer Engineering, UCLA}
\affil[2]{Department of Surgery, UCLA}
\affil[3]{Department of Bioengineering, UCLA}
\affil[4]{Medical Informatics, UCLA}
\affil[5]{Department of Surgery, UT Southwestern Medical Center, Dallas, Texas}
\affil[6]{Department of Pathology and Laboratory Medicine, UCLA}
\affil[7]{Department of Radiological Sciences, UCLA \authorcr Email: \texttt{speier@ucla.edu}}

\date{}

\renewcommand{\headeright}{}
\renewcommand{\undertitle}{}

\begin{document}

\maketitle

\setstretch{1.2}
\begin{abstract} 
\textbf{Objective:} Molecular testing (MT) classifies cytologically indeterminate thyroid nodules as benign or malignant with high sensitivity but low positive predictive value (PPV), only using molecular profiles, ignoring ultrasound (US) imaging and biopsy. We address this limitation by applying attention multiple instance learning (AMIL) to US images.\\
\textbf{Methods:} We retrospectively reviewed 333 patients with indeterminate thyroid nodules at UCLA medical center (259 benign, 74 malignant). A multi-modal deep learning AMIL model was developed, combining US images and MT to classify the nodules as benign or malignant and enhance the malignancy risk stratification of MT.\\
\textbf{Results:} The final AMIL model matched MT sensitivity (0.946) while significantly improving PPV (0.477 vs 0.448 for MT alone), indicating fewer false positives while maintaining high sensitivity.\\
\textbf{Conclusion:} Our approach reduces false positives compared to MT while maintaining the same ability to identify positive cases, potentially reducing unnecessary benign thyroid resections in patients with indeterminate nodules.
\end{abstract}

\keywords{Ultrasound, Attention, Multiple Instance Learning, Thyroid Cancer, Indeterminate Nodules.}

\setstretch{1.5}

\section{Introduction} 
\begin{figure}[ht]
\includegraphics[width=\textwidth]{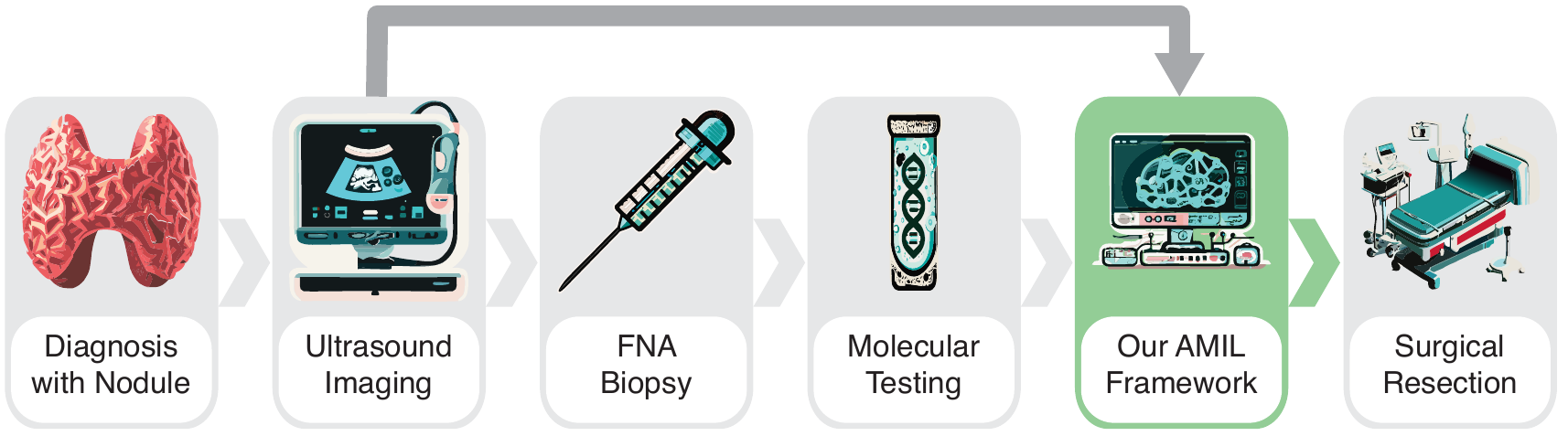}
\caption{Overview of the thyroid nodule diagnostic pipeline. Our method works as an additional step after molecular testing to combine molecular results with imaging to reduce unnecessary diagnostic surgeries.} 
\label{fig:workflow}
\end{figure}

Thyroid cancer often presents as a palpable nodule in the lower neck~\cite{RN3}. However, the estimated presence of thyroid nodules in the general population without known thyroid disease may be as high as $50-60\%$, with around $90\%$ of nodules being clinically insignificant or benign, and $7-15\%$ of cases representing thyroid cancer~\cite{RN27,RN21,RN9,RN6}. The identification of malignant thyroid nodules is critical to prevent morbidity associated with a delayed diagnosis of cancer~\cite{RN21}. Ultrasound (US) is the initial imaging modality recommended for evaluation of thyroid nodules due to its high sensitivity in identifying  high-risk nodule features such as irregular margins, punctate echogenic foci, and extrathyroidal extension~\cite{RN21,RN4}. High risk nodules on US typically warrant a fine needle aspiration (FNA), from which results are reported using the Bethesda criteria which grades nodules as I: non-diagnostic, II: benign, III: atypia of undetermined significance,  IV: follicular neoplasm, V: suspicious for malignancy, and VI: malignant~\cite{RN28,RN5,RN7}. Bethesda III and IV lesions are considered indeterminate and typically require repeat biopsy or diagnostic surgery. Molecular testing (MT) is an adjunct diagnostic tool that performs next generation sequencing on the FNA sample to assess for genetic variations, which allows for a more precise determination of malignancy risk in patients with indeterminate nodules~\cite{RN8}. MT achieved a sensitivity of $97-100\%$ and positive predictive value (PPV) of $53-63\%$ in determining the presence of malignancy in a randomized controlled trial~\cite{RN24}. While MT has proven to be useful in stratifying malignancy risk, it is calibrated to maximize sensitivity while tolerating a relatively high number of false positive results. These patients often must progress to surgery and thus a diagnostic support tool to reduce false positives while maintaining the high degree of sensitivity of MT is needed. Additionally, MT relies exclusively on sequencing, ignoring other information such as imaging, cytology, and demographics. In this work we aimed to create a multi-modal modeling pipeline to improve the ability of MT to stratify malignancy risk in patients with indeterminate thyroid nodules. We illustrate the clinical workflow with our contribution in Fig.~\ref{fig:workflow}.

Computer-aided diagnosis can significantly contribute to making informed decisions when evaluating thyroid nodule abnormalities. Previous work has shown the ability of convolutional neural networks to segment thyroid nodules from US images and classify them as benign or malignant~\cite{RN14,RN15,RN18,RN16,RN13,RN17}. Recently, Zhuang et al. demonstrated that attention multiple instance learning (AMIL) models, when applied to US images, can effectively classify patients with thyroid nodules as benign or malignant~\cite{RN25}. This work focused on malignancy prediction in patients with benign or malignant cytology on FNA, excluding those with indeterminate nodules. In this study, we specifically aimed to develop a multi-modal AMIL based approach for classifying indeterminate nodules as malignant or benign, with the goal of combining analysis of US images with MT to reduce the number of false positive results from MT while preserving sensitivity. This approach could be used to bolster the use of MT for the determination of which patients with indeterminate nodules warrant further invasive diagnostic procedures.

\section{Methods}

\subsection{Dataset} 
Patients with indeterminate thyroid nodules at UCLA Medical Center from 5/06/2016 to 2/15/2022 were retrospectively reviewed. This study was approved by the UCLA Institutional Review Board (IRB\#19-001535). Our analysis included patients with indeterminate nodules (Bethesda III and IV on FNA) who received MT. Thyroseq v3 Genomic Classifier (Sonic Healthcare, Rye Brook, NY) and Afirma GEC (Veracyte, San Francisco, CA) were used for MT. For those patients who underwent surgery, the surgical pathology was used to determine a label of benign or malignant. In those who did not undergo surgery, benign results were assumed as these patients were determined to be at low risk of nodule malignancy based on clinical evaluation including a benign MT result, an approach that has been used in previous studies~\cite{RN24}. A significant number of patients had a pathologic diagnosis of non-invasive follicular thyroid neoplasm with papillary like nuclear features (NIFTP). These patients were grouped with those diagnosed with thyroid cancer as NIFTP is currently considered pre-malignant and should be managed surgically~\cite{RN26,RN24}. Complete demographic comparisons between patients with benign and malignant thyroid nodules is tabulated in Table~\ref{tab:demographic_comparisons}.

\begin{table}[ht]
\centering
\caption{Demographic comparisons between patients with benign and malignant thyroid nodules. IQR: interquartile range; MT: molecular testing; NIFTP: non-invasive follicular thyroid neoplasm with papillary like nuclear features; FNA: fine-needle aspiration.}
\label{tab:demographic_comparisons}
\begin{tabular*}{0.7\textwidth}{@{\extracolsep{\fill}}|l@{\hspace{2em}}|r|r|@{\extracolsep{\fill}}}
\hline
\textbf{Variable} & \textbf{Benign} $(n=259)$ & \textbf{Malignant} $(n=74)$ \\
\hline
Gender = Male (\%) & $49\,(19.1)$ & $16\,(21.1)$ \\
Age (median [IQR]) & $56\,[43,68]$ & $47\,[37,60]$ \\
Surgical Resection (\%) & 101 (39.0) & $74\,(100.0)$ \\
Bethesda IV (\%) & 42 (16.3) & 20 (26.3) \\
MT Suspicious (\%) & 86 (33.5) & 71 (93.4) \\
\hline
Surgical Pathology (\%) & & \\
\quad Benign & $101\,(39.0)$ & $0\,(00.0)$ \\
\quad Papillary Thyroid Cancer & $0\,(0.0)$ & $35\,(47.3)$ \\
\quad Follicular Carcinoma & $0\,(0.0)$ & $11\,(14.9)$ \\
\quad NIFTP & $0\,(0.0)$ & $26\,(35.1)$ \\
\quad Other Thyroid Cancer & $0\,(0.0)$ & $2\,(02.7)$ \\
\hline
\end{tabular*}
\end{table}

\subsection{Attention Multiple Instance Learning} 
The Multiple Instance Learning (MIL)~\cite{Maron_1997} framework is a form of weakly supervised learning where labels are provided for groups of instances, or bags, rather than individual instances. In our study, this is relevant as we deal with multiple US scans per patient, without specific annotations for each frame or patch. We treat all scans of a patient as a single bag, assigning a malignancy label based on the patient's surgical outcome, allowing the model to predict outcomes without detailed scan-level annotations.

A limitation of traditional MIL is its inability to specify which instances contribute to the bag’s label. Attention-based MIL~\cite{ilse_attention-based_2018} (AMIL) addresses this by providing attention scores that indicate each scan's contribution to the overall label. We use a gated attention mechanism with $tanh$ and $sigmoid\,(\sigma)$ activations, introducing a learnable non-linearity to improve the model’s ability to capture complex relationships. The AMIL module is defined as:
\begin{equation}
    \mathbf{z} = \left[softmax([\sigma(\mathbf{h}\cdot W_f) \odot \tanh(\mathbf{h}\cdot W_g)]\mathbf{b})^\top \mathbf{h} \right] \mathbf{c}
\end{equation}
where $\mathbf{h} \in \mathbb{R}^{k \times 256}$ is the US image feature embeddings, $W_f,\, W_g \in \mathbb{R}^{256 \times 128}$ are learnable weights of the gated attention mechanism, $\mathbf{b} \in \mathbb{R}^{128 \times 1}$ and $\mathbf{c} \in \mathbb{R}^{256 \times 16}$ are learnable linear transformations, $\mathbf{z} \in \mathbb{R}^{1 \times 16}$ is the AMIL output, and $\odot$ is elementwise multiplication. To reduce model complexity, the feature extractor output $\mathbf{H} \in \mathbb{R}^{k \times 2048}$ is linearly transformed to $\mathbf{h} \in \mathbb{R}^{k \times 256}$ before passing to AMIL. AMIL enhances explainability, providing attention scores that highlight influential regions in US images, offering clinicians interpretable evidence to inform diagnostics and treatment.

\subsection{Our Framework} 
During preprocessing, US images were cropped to eliminate extraneous elements like peripheral text and probe orientation markers. For the whole frame configuration, scans were resized to standardize the larger dimension to 256 pixels, with square padding applied to the shorter dimension for uniformity across the dataset. To address the high computational cost of pixel-level image processing, we also adopted a patch-based strategy, extracting $256 \times 256$ and $128 \times 128$ patches, the latter resized to $256 \times 256$ for compatibility with our pre-trained feature extractor. This method, employing custom strides, focuses on localized features without extra padding or cropping, followed by normalization using training set statistics and duplication of grayscale images across three color channels for the extractor's requirements. Following preprocessing, we employ a ResNet101 model~\cite{He2016} pre-trained on ImageNet to extract features from each image or patch. Using a pre-trained backbone is advantageous, as it leverages deep learning models trained on large datasets, providing robust feature representations without fine-tuning on our specific medical data. 

\begin{figure}[t]
\includegraphics[width=\textwidth]{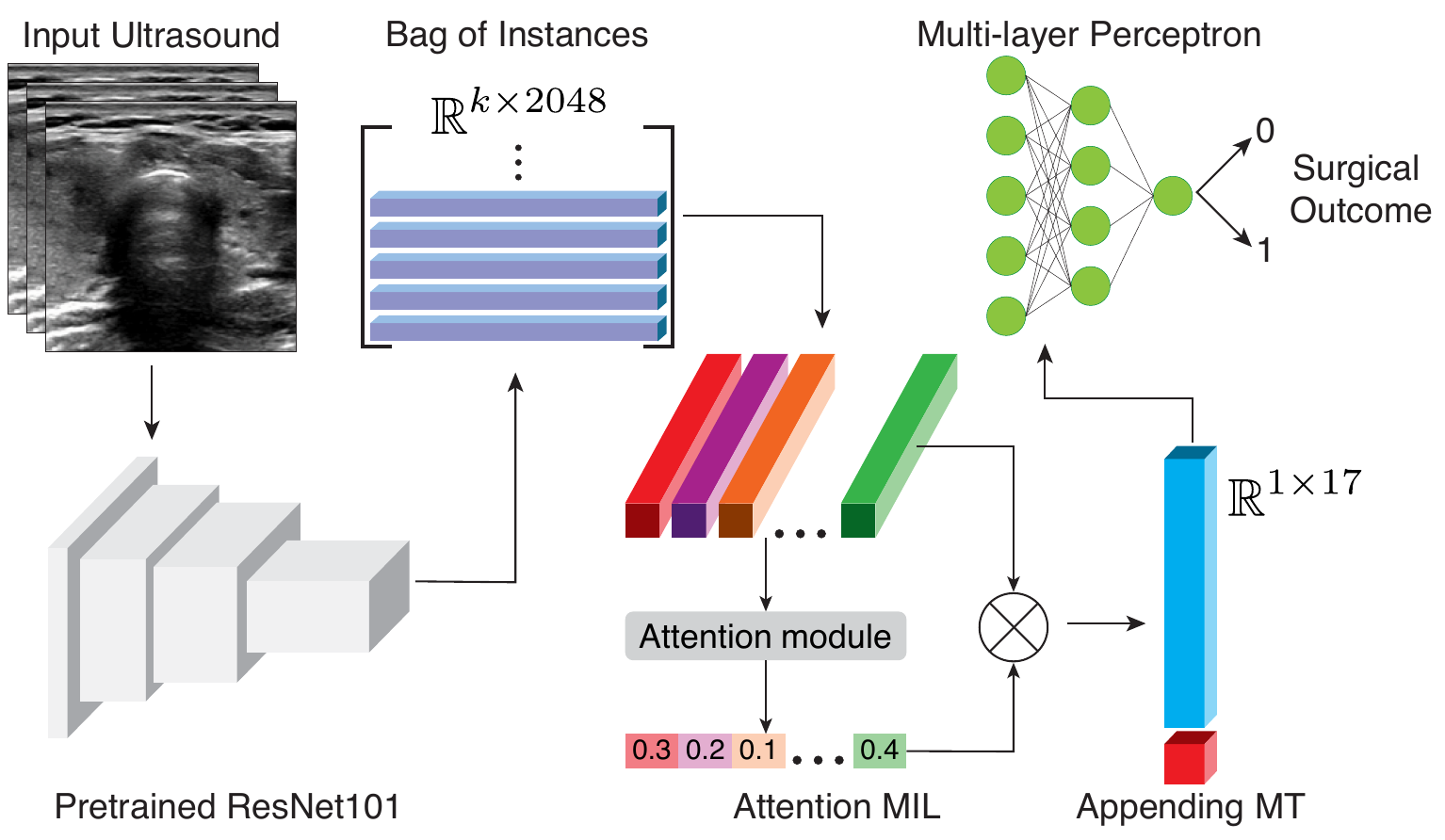}
\caption{An end-to-end overview of the multi-modal AMIL framework.} \label{fig:architecture}
\end{figure}

From each US image ($\mathbf{X}_i \in \mathbb{R}^{3 \times 256 \times 256}$), we derive a feature vector $\mathbf{H}_i \in \mathbb{R}^{1 \times 2048}$ from ResNet101’s final hidden layer. For a patient with $k$ US scans, feature vectors are stacked to form the feature matrix $\mathbf{H} \in \mathbb{R}^{k \times 2048}$ when using whole frames. With $256 \times 256$ or $128 \times 128$ patches, having $n$ and $m$ patches per scan, matrices $\mathbf{H} \in \mathbb{R}^{nk \times 2048}$ and $\mathbf{H} \in \mathbb{R}^{mk \times 2048}$ are generated respectively. These are processed through AMIL~\cite{ilse_attention-based_2018}, which assigns attention weights to produce an attention-weighted feature vector $\mathbf{z} \in \mathbb{R}^{1 \times 16}$ encoding all patient scan information. This vector is concatenated with the binary MT outcome to form $\mathbf{z'} \in \mathbb{R}^{1 \times 17}$, which feeds into a fully connected network to predict the final surgical outcome. The framework is illustrated in Fig.~\ref{fig:architecture}.

\subsection{Experiments} 
We conducted baseline experiments to evaluate the predictive power of clinical features on surgical outcomes for thyroid nodules, using binary MT outcomes as the initial benchmark, given MT's status as the clinical gold standard~\cite{RN8,RN24}. To enhance our predictive model, we integrated binarized Bethesda III and IV scores with binary MT outcomes (MT+BE), providing a more nuanced classification that incorporates both genetic and cytological information. However, since the optimal decision threshold in the MT+BE model ended up relying solely on MT, completely ignoring the Bethesda scores, we proceeded with MT alone for all further experiments involving US images. Next, we obtained US image features by extracting one feature vector $\mathbf{z} \in \mathbb{R}^{1 \times 16}$ for each patient from their US scans ($\mathbf{H} \in \mathbb{R}^{k \times 2048}$) using AMIL. This vector was then concatenated with clinical features to predict surgical outcomes. The methodology was tested across various image resolutions, including whole-frame images and patches of sizes $256 \times 256$ and $128 \times 128$, to determine the most effective image analysis granularity. Finally, the three configurations were ensembled during test time to produce our final model configuration. The proposed models' statistical significance relative to the baseline MT were determined using a one-sided Wilcoxon signed rank test, at a significance level of 0.05. We trained our models using a Weighted Binary Cross-Entropy loss function to address the class imbalance by assigning higher weight to the minority positive class. The function is expressed as:
\begin{equation}
    BCE_w(y,\,p) = - \alpha \, y \log(p) - (1 - y) \log(1 - p)
\end{equation}
where $y$ is the ground truth, $p$ is the predicted probability, and $\alpha$ is the minority class weight. This approach is crucial given the class imbalance and the importance of identifying hard-to-classify samples. We used the AdamW optimizer, known for handling sparse gradients and weight decay~\cite{loshchilov_decoupled_2019}. The learning rate, set to 0.005, was dynamically adjusted using a ReduceLROnPlateau scheduler which halved the rate upon validation F1 score plateaus, ensuring adaptive learning~\cite{mukherjee_simple_2019}. Training was conducted over 100 epochs with a batch size of one, incorporating five-fold cross-validation, stratified by surgical outcomes. For all configurations, we did a 90/10 split on the training fold to create the validation set. Models were implemented in PyTorch; code is available \href{https://github.com/ShreeramAthreya/ThyroidIndetClassify}{here}.

\begin{table}[!ht]
\centering
\caption{Model performance. Mean and standard deviation of metrics across five folds MT: Molecular Testing, BE: Bethesda scores.}
\label{tab:results}
\begin{tabular*}{0.9\textwidth}{@{\extracolsep{\fill}}|l|*{6}{r|}@{\extracolsep{\fill}}}
\hline
Models                           &       AUROC  &     Accuracy &  Sensitivity &  Specificity &          PPV &     F1 Score \\ 
\hline
\multirow{2}{*}{MT + BE}         &        0.728 &        0.727 &\textbf{0.946}&        0.664 &        0.448 &        0.607 \\
                                 &  $\pm 0.037$ &  $\pm 0.032$ &  $\pm 0.050$ &  $\pm 0.046$ &  $\pm 0.024$ &  $\pm 0.025$ \\
\hline
\multirow{2}{*}{MT + Whole Frame}&        0.828 &        0.733 &\textbf{0.946}&        0.671 &        0.456 &        0.614 \\
                                 &  $\pm 0.053$ &  $\pm 0.044$ &  $\pm 0.050$ &  $\pm 0.063$ &  $\pm 0.039$ &  $\pm 0.034$ \\
\hline
\multirow{2}{*}{MT + Patch 128}  &        0.819 &        0.742 &\textbf{0.946}&        0.683 &        0.464 &        0.621 \\
                                 &  $\pm 0.029$ &  $\pm 0.034$ &  $\pm 0.050$ &  $\pm 0.058$ &  $\pm 0.033$ &  $\pm 0.020$ \\
\hline
\multirow{2}{*}{MT + Patch 256}  &        0.804 &       0.733  &\textbf{0.946}&        0.671 &        0.453 &        0.612 \\
                                 &  $\pm 0.028$ &  $\pm 0.033$ &  $\pm 0.050$ &  $\pm 0.045$ &  $\pm 0.027$ &  $\pm 0.029$ \\
\hline
\multirow{2}{*}{Ensemble}        &\textbf{0.831}&\textbf{0.757}&\textbf{0.946}&\textbf{0.703}&\textbf{0.477}&\textbf{0.634}\\
                                 &  $\pm 0.041$ &  $\pm 0.017$ &  $\pm 0.050$ &  $\pm 0.034$ &  $\pm 0.016$ &  $\pm 0.010$ \\
\hline
\end{tabular*}
\end{table}

\begin{figure}[!htb]
\includegraphics[width=\textwidth]{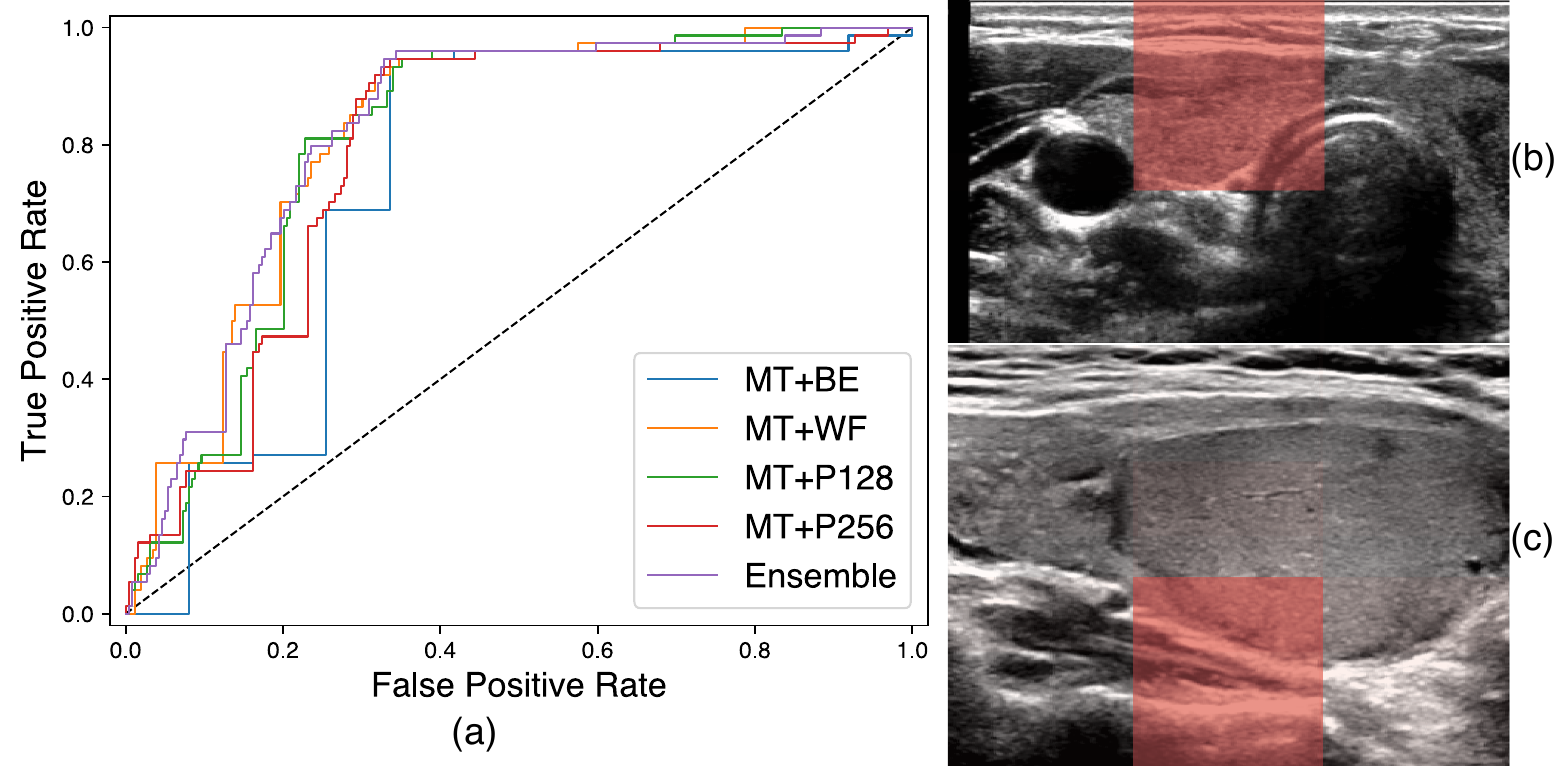}
\caption{Results of the AMIL framework. (a) ROC curves comparing different model configurations, (b) and (c) Patch 256 attention maps of representative US scans. In (b), there is a nodule in the right lobe of the thyroid with the attention map including the majority of the nodule. The deep border has a notable rim of hyperechogenicity that is partially contained within the attention map. In (c), there is a large thyroid nodule in the center of the frame, containing a central solid component with hypoechoic peripheral areas. Notably, the attention map is located directly over an area containing isoechoic and hypoechoic nodule components at the periphery as well as the hyperechoic interface between the nodule and the deep surrounding tissue. BE: Bethesda scores, WF: Whole frame images.} 
\label{fig:results}
\end{figure}

\section{Results} 
There were 333 patients with indeterminate thyroid nodules who met this study's inclusion criteria (259 benign and 74 malignant). The performance of the different configurations is presented in Table~\ref{tab:results}. The models' predictions aggregated over all 5 folds were used to plot the ROC curves shown in Fig.~\ref{fig:results}(a). The baseline MT+BE model achieved an AUROC of 0.728 with a sensitivity of 0.946, consistent with prior studies~\cite{RN24}. Incorporating whole-frame US scans improved the AUROC to 0.828, with marginal gains in specificity and PPV. The Patch 128 model achieved an AUROC of 0.819, with notable improvements in accuracy (0.742), specificity (0.683), and F1 Score (0.621). The Patch 256 configuration demonstrated an AUROC of 0.804, maintaining high sensitivity (0.946) but showing slightly lower specificity and PPV compared to Patch 128. The Ensemble model outperformed all others, achieving the highest AUROC of 0.831 and PPV of 0.477, with statistically significant improvements $(p=0.0008)$, indicating its superior ability to reduce false positives while maintaining high sensitivity.

By selecting a decision threshold with sensitivity at least matching MT, our model aimed to reduce unnecessary surgeries by identifying patients incorrectly classified as malignant by MT. Though the improvement in PPV is modest (0.477 vs. 0.448), this could lead to a significant clinical impact. Given over 120,000 indeterminate biopsies annually in the U.S.~\cite{Sosa2013,RN21}, even a small reduction in false positive rate could lead to a significant number of avoided surgeries. Moreover, providing continuous output rather than binary predictions allows physicians more flexibility in surgical decisions (e.g., thyroidectomy, lobectomy, or no surgery). Patch attention map visualizations in Fig.~\ref{fig:results}(b) and (c) show that the model focuses on clinically relevant portions of the nodules when making malignancy predictions, enhancing the interpretability of the AMIL framework's decision-making process.

When combining the Bethesda score with MT, the AUROC increased from 0.627 (MT alone) to 0.728, indicating biopsy results provide complementary information to MT. Using the Bethesda score in combination with MT did not show a significant improvement in other metrics, as the optimal decision point primarily relied on the MT output. Future work could include digitized cytology slides, which contain more information than the Bethesda score. However, the current model holds significant clinical value, as cytology slides are not routinely digitized due to the time and cost associated with scanning.

\section{Conclusion} 
We developed an AMIL model that combines US images and MT to classify indeterminate nodule malignancy, which retains MT's high sensitivity while significantly reducing false positive results. In thyroid cancer diagnostics, minimizing false negatives is crucial to prevent delays in diagnosis and treatment for cancer patients, while false positive results can lead to unnecessary diagnostic surgery with incurred cost and associated risk of morbidity. Our model reduces the number of false positive results while preserving the same degree of sensitivity offered by MT, potentially decreasing unnecessary diagnostic surgeries without an increased risk of misdiagnosis of thyroid cancer. Future work could involve validating our approach in a larger prospective study prior to clinical translation.

\section{Acknowledgements and Funding}
This work was supported by the National Institute of Biomedical Imaging and Bioengineering of the National Institutes of Health under award number R21EB030691 and a UCLA Radiology Exploratory Research Grant.

\section{Author Contributions}
SA contributed to conceptualization, formal analysis, methodology, investigation, visualization, and writing the original draft. AM and SSAS were responsible for software development, methodology, investigation, validation, and writing the original draft. VI and AR handled data curation, methodology, and writing review and editing. VS, CM, HZ, MP, and RM contributed to data curation, resources, validation, and writing review and editing. CA and WS provided funding acquisition, supervision, project administration, and writing review and editing.

\section{Conflict of Interest}
None declared.

\section{Data Availability}
The data underlying this article will be shared on reasonable
request to the corresponding author.

\bibliographystyle{ama.bst}
\bibliography{reference}

\end{document}